# Vapor-Cell-Induced Uncertainty in Rydberg Atom Measurements via the Electric-Field Volume-Integral-Equation Method

Martin Štumpf, *Senior Member, IEEE*, William J. Watterson, Rajavardhan Talashila, Matt T. Simons, Alexandra Artusio-Glimpse, Lawrence Carslake, Tian Hong Loh, *Fellow, IEEE*, and Christopher L. Holloway, *Fellow, IEEE*

***Abstract*—Electromagnetic scattering effects of a vapor cell on electric-field measurements using Rydberg atom-based sensors are analyzed with the aid of the volume integral equation method. In a manner similar to measurement, this computational approach determines the electric field over grid points within the vapor cell. Its relatively high computational efficiency makes it suitable for use in optimization routines and statistical uncertainty studies. We apply this method to compare uncertainty contributions arising due to the presence of the vapor cell, such as uncertainty in the glass relative permittivity or standing wave formation inside the cell, to those arising from the atomic spectroscopic measurement, such as uncertainty in the atomic dipole moment. For vapor cell dimensions less than half a wavelength, the dominant uncertainty source arises from uncertainty in the glass relative permittivity, resulting in a total uncertainty of $\sim$3.5% – comparable to the best uncertainties obtained with traditional field generation methods at national metrology institutes. Precise permittivity measurements have the potential to further reduce measurement uncertainty to $< 1\%$.**

## I. Introduction

RYDBERG-atom-based spectroscopy is an emerging experimental technique for SI-traceable, self-calibrated and highly sensitive electric ($E$)-field measurements [1]–[4]. This method optically excites an atomic vapor inside a glass vapor cell to a high principal quantum number "Rydberg" state ($n \gtrsim 20$), where the large induced atomic dipole moments cause the atoms to become highly sensitive to external electric fields. Although the atoms can provide direct SI-traceability with low uncertainty of the electric field strength *inside* the glass vapor cell, a key consideration for $E$-field strength metrology is to measure the incident field strength *outside* the vapor cell. Thus, understanding the electromagnetic (EM) scattering and absorption properties of vapor cells is essential to obtaining a reliable assessment of the measurement uncertainty.

The geometry of a vapor cell can significantly influence the accuracy of $E$-field measurements obtained with Rydberg atom-based sensors. In fact, the vapor cell acts as an open resonator, and when illuminated by external EM fields, it can support the formation of non-uniform, standing-wave EM fields. This behavior was investigated in Ref. [5], for example, both experimentally and through a simple one-dimensional Fabry-Pérot resonator model. The study in [5] concludes that the vapor cell should be kept relatively small compared with the wavelength in order to minimize the occurrence of non-uniform internal EM field patterns.

To compute the detailed $E$-field distribution inside a vapor cell, commercial EM-field solvers based on the finite-element method (FEM) are most frequently employed (e.g., [6]–[8]). These computational tools, however, are not open-source and, consequently, their implementation details are not publicly disclosed. As a result, they do not provide the fully controlled computational environment required to reliably assess the uncertainties associated with the EM-scattering properties of vapor cells. To address this limitation, we implemented a relatively simple computational framework that enables systematic statistical uncertainty studies. More specifically, in this paper we analyze the effect of a vapor cell on the $E$-field sensing with the aid of the method of moments (MoM) [9] applied to the electric-field volume integral equation (VIE) [10], [11]. The presented MoM-VIE approach has several advantages over other numerical studies based on finite-difference/FEM methodologies: (a) its spatial computational domain is limited to the volume occupied by a vapor cell only. As a result, it is unnecessary to discretize the surrounding space and introduce "matched layers" to terminate the solution domain – the radiation condition is implicitly incorporated in the EM Green's tensor. Moreover, (b) the presented MoM-VIE approach computes only the $E$-field distribution inside a vapor cell – no other EM field quantities are calculated and/or stored during the computation. Therefore, this approach is relatively computationally efficient and reflects the actual experimental

(*Corresponding author: Martin Štumpf*) M. Štumpf was with the National Institute of Standards and Technology (NIST), 325 Broadway Boulder, CO 80305, USA, on sabbatical leave from the Lerch Laboratory of EM Research, Department of Radio Electronics, FEEC, Brno University of Technology, Technická 3082/12, 616 00 Brno, The Czech Republic and from the Department of Computer Science, Electrical and Space Engineering, EISLAB, Luleå University of Technology, 971 87 Luleå, Sweden (e-mail: martin.stumpf@centrum.cz).

L. Carslake and T. H. Loh are with the National Physical Laboratory (NPL), Teddington, UK.

C. L. Holloway, W. J. Watterson, M. T. Simons, A. Artusio-Glimpse and R. Talashila are with the National Institute of Standards and Technology (NIST), 325 Broadway Boulder, CO 80305, USA (e-mail: christopher.holloway@nist.gov).

procedure involving Rydberg atom-based $E$-field sensors.

To demonstrate the utility of this MoM-VIE method, we determine the Rydberg sensor measurement uncertainty. The *Guide to the expression of uncertainty in measurement* outlines two methods for evaluating uncertainty – Type A and Type B [12]. Type A uncertainties are statistically determined from a number of repeated observations, either experimentally or computationally, while, Type B uncertainties are estimated from other means and usually involve an educated operator's best estimates on the limits and distribution of expected variances. In most cases, a Type A evaluation is preferable to Type B but comes at the expense of time and resources to evaluate completely. A primary motivation for creating the efficient MoM-VIE solver presented here, was to enable a Type A uncertainty evaluation through a Monte Carlo repeated random sampling method using a non-commercial EM solver to computationally determine the uncertainties associated with variations in vapor cell parameters such as dielectric constant. We apply this Type A analysis to uncertainty components related to the dielectric constant and angle of incidence. Due to finite discretization of the VIE mesh, two of our uncertainty sources, related to beam position in the glass vapor cell, however, are more amenable to a Type B analysis.

The paper is organized as follows. Section II defines the EM scattering problem associated with a vapor cell. The corresponding VIE is formulated in Section III. This section is supplemented with Appendix A, where selected properties of its Green's kernel are discussed. A straightforward numerical solution of the VIE is presented in Section IV. Further, Sec. V provides illustrative numerical examples, including a validation study and an extensive analysis of both Type A and Type B uncertainties for a selected vapor cell. Section VI gives a discussion on applications of the presented methodology. Finally, conclusions are drawn in Sec. VII.

## II. Problem Definition

We analyze an EM scattering problem shown Fig. 1. It consists of a vapor cell located in vacuum that is irradiated by an EM incident wave. The EM properties of the ambient medium in $\mathcal{D}^\infty$ are described by (scalar, real-valued and positive) electric permittivity $\epsilon_0$ and magnetic permeability $\mu_0$. The corresponding EM wave speed is $c_0 = (\epsilon_0\mu_0)^{-1/2} > 0$. Our EM model of the vapor cell is represented by an EM-penetrable resonator with an interior region, $\mathcal{D}_0$, filled by vacuum. Its walls, occupying a bounded domain $\mathcal{D}_1$, differ with respect to the surrounding medium in its electric permittivity $\hat{\epsilon}_1$ and electric conductivity $\hat{\sigma}_1$. Both $\hat{\epsilon}_1$ and $\hat{\sigma}_1$ can, in principle, be spatially dependent and dispersive. The total domain occupied by the vapor cell is denoted by $\mathcal{D} = \mathcal{D}_0 \cup \mathcal{D}_1$. The analysis is carried out in the (one-sided) Laplace-transform domain with $\{s \in \mathbb{C}; \mathrm{Re}(s) > 0\}$ being its parameter (= complex frequency) [13, Section B.1].

The position in the problem configuration is localized using coordinates $\{x_1, x_2, x_3\}$ defined with respect to a Cartesian reference with its origin $\mathcal{O}$ and the standard base $\{\boldsymbol{i}_1, \boldsymbol{i}_2, \boldsymbol{i}_3\}$. Consequently, the position vector can be specified as $\boldsymbol{x} = x_1\boldsymbol{i}_1 + x_2\boldsymbol{i}_2 + x_3\boldsymbol{i}_3$. Partial differentiation with respect to $x_m$ is then denoted by $\partial_m$. The Einstein summation convention over repeated subscripts is employed [13, Sec. A.2]. For example, if $a_k$ and $b_k$ with $k \in \{1, 2, 3\}$ denote one-dimensional arrays, then $a_k b_k$ stands for $a_1b_1 + a_2b_2 + a_3b_3$.

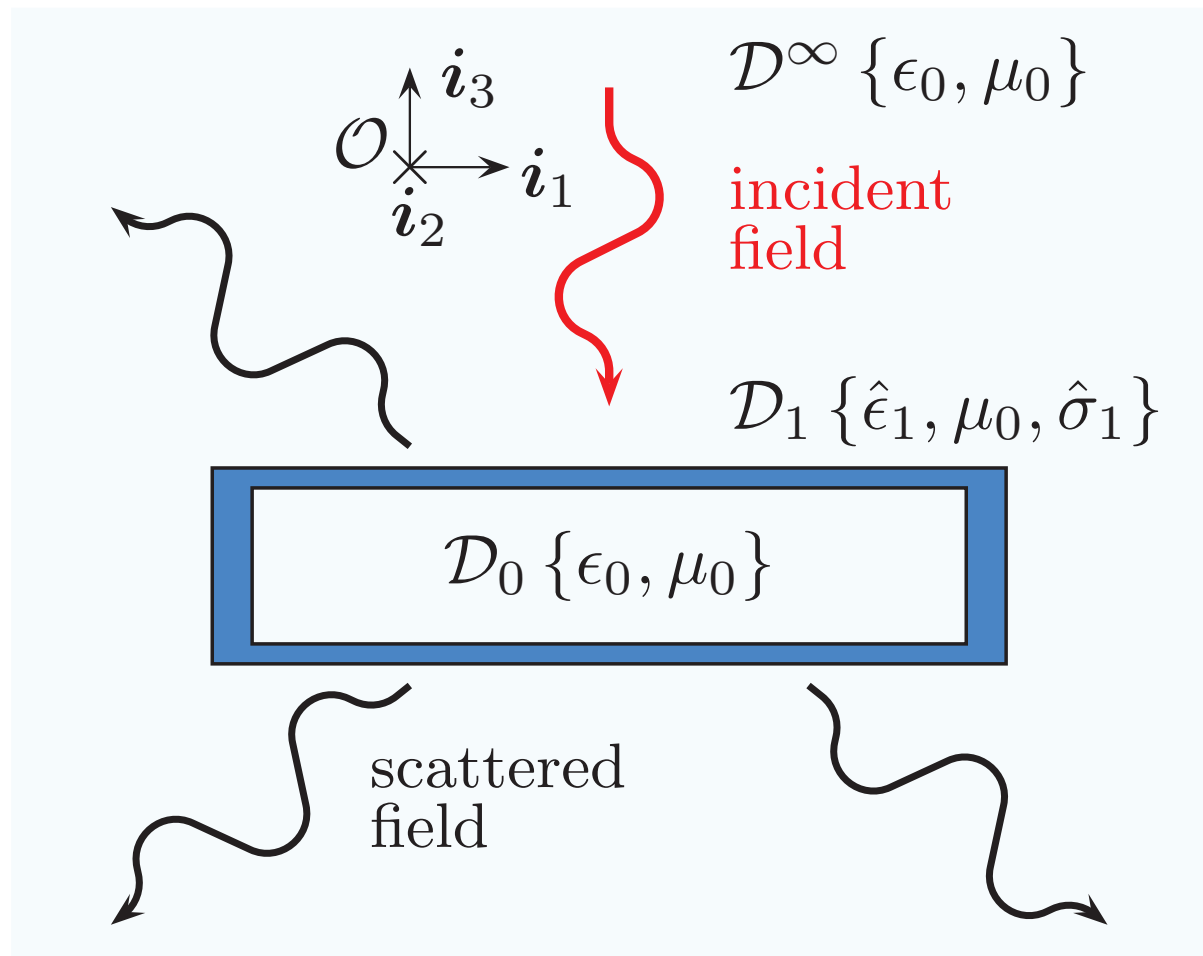

Fig. 1. Cross-section of vapor cell in the presence of incident EM field.

The ($k$th component of the) incident electric-field strength (i.e., the electric field that would be present in the configuration if the vapor cell showed no contrast with respect to the background medium) is a function of the spatial position and of the parameter $s$, and is denoted by $\hat{E}_k^{\mathrm{i}} = \hat{E}_k^{\mathrm{i}}(\boldsymbol{x}, s)$. EM scattering effects due to the vapor cell are accounted for by the scattered electric-field strength (denoted by superscript $^{\mathrm{s}}$) that is thus defined as the difference between the total electric field and the incident one, i.e.,

$$\hat{E}_k^{\mathrm{s}}(\boldsymbol{x}, s) = \hat{E}_k(\boldsymbol{x}, s) - \hat{E}_k^{\mathrm{i}}(\boldsymbol{x}, s), \tag{1}$$

for all $k \in \{1, 2, 3\}$. The analysis that follows yields the total electric field at grid points in $\mathcal{D}$, i.e., inside the vapor cell and its EM-penetrable walls.

## III. Problem Formulation

The problem is analyzed with the aid of a source-type integral representation for the scattered electric field [13, (28.9-43)]

$$\hat{E}_r^{\mathrm{s}}(\boldsymbol{x}, s) = \int_{\boldsymbol{x}' \in \mathcal{D}} \hat{G}_{r,k}^{EJ}(\boldsymbol{x} - \boldsymbol{x}', s) \hat{J}_k^{\mathrm{c}}(\boldsymbol{x}', s) \mathrm{d}V', \tag{2}$$

for all $r \in \{1, 2, 3\}$, where $\hat{G}_{r,k}^{EJ}$ is the electric-field/electric-current Green's tensor pertaining to the embedding [13, Sec. 28.8] and $\hat{J}_k^{\mathrm{c}}$ is the equivalent contrast volume source density of electric current

$$\hat{J}_k^{\mathrm{c}}(\boldsymbol{x}, s) = [\hat{\eta}_1(\boldsymbol{x}, s) - \hat{\eta}_0(s)] \hat{E}_k(\boldsymbol{x}, s), \tag{3}$$

where $\hat{\eta}_1 = s\hat{\epsilon}_1 + \hat{\sigma}_1$ and $\hat{\eta}_0 = s\epsilon_0$. Using $\hat{G}_{r,k} = \hat{\eta}_0 \hat{G}_{r,k}^{EJ}$, the integral representation can be rewritten as

$$\hat{E}_r^{\mathrm{s}}(\boldsymbol{x}, s) = \int_{\boldsymbol{x}' \in \mathcal{D}} \hat{G}_{r,k}(\boldsymbol{x} - \boldsymbol{x}', s) \hat{\chi}(\boldsymbol{x}', s) \hat{E}_k(\boldsymbol{x}', s) \mathrm{d}V', \tag{4}$$

where $\hat{\chi} = \hat{\eta}_1/\hat{\eta}_0 - 1$ represents the *relative* electric-contrast function and Green's tensor can be expressed as (cf., [13, p. 851])

$$\hat{G}_{r,k}(\boldsymbol{x}-\boldsymbol{x}',s) = -(s^2/c_0^2)\hat{g}(\boldsymbol{x}-\boldsymbol{x}',s)\delta_{r,k} + \partial_r\partial_k\hat{g}(\boldsymbol{x}-\boldsymbol{x}',s), \quad (5)$$

where

$$\hat{g}(\boldsymbol{x}-\boldsymbol{x}',s) = \frac{\exp[-s|\boldsymbol{x}-\boldsymbol{x}'|/c_0]}{4\pi|\boldsymbol{x}-\boldsymbol{x}'|} \quad \text{for } \boldsymbol{x} \neq \boldsymbol{x}', \quad (6)$$

is the scalar Green's function that is for $\mathrm{Re}(s) > 0$ bounded as $|\boldsymbol{x}-\boldsymbol{x}'| \to \infty$.

For the observation point inside the vapor cell, the use of (1) in (4) leads to an electric-field VIE. Owing to the strong singularity of $\hat{G}_{r,k}$, however, the integral in (4) for $\boldsymbol{x} \in \mathcal{D}$ has to be interpreted in the sense of Cauchy principal value [14], i.e.,

$$\hat{E}_r(\boldsymbol{x},s)\left[1+\tfrac{1}{3}\hat{\chi}(\boldsymbol{x},s)\right] - \text{P.V.}\int_{\boldsymbol{x}'\in\mathcal{D}} \hat{G}_{r,k}(\boldsymbol{x}-\boldsymbol{x}',s)\hat{\chi}(\boldsymbol{x}',s)\hat{E}_k(\boldsymbol{x}',s)\mathrm{d}V' = \hat{E}^{\mathrm{i}}_r(\boldsymbol{x},s) \quad \text{for } \boldsymbol{x} \in \mathcal{D}. \quad (7)$$

where the spherical "principal volume" (P.V.) has been assumed (see [15] and Appendix A for more details). Integral equation (7) is further solved numerically.

## IV. Numerical Solution

To solve (7) numerically, the spatial domain $\mathcal{D} = \mathcal{D}_0 \cup \mathcal{D}_1$ of the vapor cell (see Fig. 1) is discretized. For the sake of simplicity, the solution domain is divided uniformly into $N$ identical cubes, i.e., $\mathcal{D} \simeq \cup_{n=1}^{N}\Delta_n$. Each elementary cube occupies $\Delta_n = \{x_1^{[n]} - h/2 < x_1 < x_1^{[n]} + h/2, x_2^{[n]} - h/2 < x_2 < x_2^{[n]} + h/2, x_3^{[n]} - h/2 < x_3 < x_3^{[n]} + h/2\}$ with $n = \{1, \cdots, N\}$, where $h > 0$ denotes its (relatively small) side length and $\boldsymbol{x}^{[n]} = x_1^{[n]}\boldsymbol{i}_1 + x_2^{[n]}\boldsymbol{i}_2 + x_3^{[n]}\boldsymbol{i}_3$ is the position of its center. Evaluating the VIE at the center of the $m$th cube located at $\boldsymbol{x} = \boldsymbol{x}^{[m]}$, we obtain

$$\hat{E}_r^{[m]}\left(1+\tfrac{1}{3}\hat{\chi}^{[m]}\right) - \sum_{\substack{n=1\\n\neq m}}^{N}\int_{\boldsymbol{x}'\in\Delta_n}\hat{G}_{r,k}(\boldsymbol{x}^{[m]}-\boldsymbol{x}',s)\hat{\chi}(\boldsymbol{x}',s)\hat{E}_k(\boldsymbol{x}',s)\mathrm{d}V' - \text{P.V.}\int_{\boldsymbol{x}'\in\Delta_m}\hat{G}_{r,k}(\boldsymbol{x}^{[m]}-\boldsymbol{x}',s)\hat{\chi}(\boldsymbol{x}',s)\hat{E}_k(\boldsymbol{x}',s)\mathrm{d}V' = \hat{E}_r^{\mathrm{i};[m]} \quad \text{for all } m \in \{1,\cdots,N\}, \quad (8)$$

where we have excluded the "diagonal" principal-value integral from the summation and used the short-hand notation, e.g., $\hat{E}_r^{[m]} = \hat{E}_r(\boldsymbol{x}_m, s)$. If we replace the $m$th cube $\Delta_m$ with an equivalent sphere of radius $h_0 = h(3/4\pi)^{1/3}$, the principal-value integral can be evaluated with the aid of (32) as

$$\text{P.V.}\int_{\boldsymbol{x}'\in\Delta_m}\hat{G}_{r,k}(\boldsymbol{x}^{[m]}-\boldsymbol{x}',s)\hat{\chi}(\boldsymbol{x}',s)\hat{E}_k(\boldsymbol{x}',s)\mathrm{d}V' = \int_{\boldsymbol{x}'\in\Delta_m\setminus\mathbb{B}_\delta}\hat{G}_{r,k}(\boldsymbol{x}^{[m]}-\boldsymbol{x}',s)\hat{\chi}(\boldsymbol{x}',s)\hat{E}_k(\boldsymbol{x}',s)\mathrm{d}V' \simeq \frac{2}{3}\left[\left(1+\frac{sh_0}{c_0}\right)\exp\left(-\frac{sh_0}{c_0}\right)-1\right]\hat{\chi}^{[m]}\hat{E}_r^{[m]}, \quad (9)$$

where $\simeq$ denotes approximation of order of $O(h^2)$. The use of (9) in (8) gives

$$\hat{E}_r^{[m]}\left\{1+\hat{\chi}^{[m]}\left[1-\frac{2}{3}\left(1+\frac{sh_0}{c_0}\right)\exp\left(-\frac{sh_0}{c_0}\right)\right]\right\} - \sum_{\substack{n=1\\n\neq m}}^{N}\hat{\chi}^{[n]}\hat{E}_k^{[n]}\int_{\boldsymbol{x}'\in\Delta_n}\hat{G}_{r,k}(\boldsymbol{x}^{[m]}-\boldsymbol{x}',s)\mathrm{d}V' \simeq \hat{E}_r^{\mathrm{i};[m]} \quad \text{for all } m \in \{1,\cdots,N\}, \quad (10)$$

The remaining integral of the Green's tensor over the $n$th cube $\Delta_n$ for $n \neq m$ can be evaluated numerically using any standard integration routine. In our implementation, the volume integral is approximated by the 3-point Gauss quadrature [16, p. 916].

Equation (10) represents a discrete version of the VIE (7) that can be cast into the following matrix form

$$\bar{\boldsymbol{G}} \cdot \boldsymbol{E} = \boldsymbol{F}, \quad (11)$$

where $\bar{\boldsymbol{G}}$ is the $[3N\times 3N]$ system array, $\boldsymbol{E}$ is an $[3N\times 1]$ array of (unknown) $\{\hat{E}_1, \hat{E}_2, \hat{E}_3\}$-field values at the grid points and $\boldsymbol{F}$ is an $[3N\times 1]$ array of (known) incident electric-field values. Finally, once $\bar{\boldsymbol{G}}$ and $\boldsymbol{F}$ are specified, the system of equations can be (iteratively) solved for the $E$-field distribution over the chosen spatial grid of the discretized vapor cell. Illustrative examples are presented in the following Section V.

## V. Illustrative Numerical Examples

The MoM-VIE method for evaluating the EM scattering effects of vapor cells has been implemented in MATLAB®[1]. This section features selected examples demonstrating the capabilities of our computational tool.

As the first example, we analyze the $E$-field distribution inside a cubic vapor cell with $10\,\mathrm{mm}$ sides [5]. The thickness of its walls made of Pyrex[1] is $1\,\mathrm{mm}$. The cell is centered at the coordinate system origin $\mathcal{O}$. The vapor cell is irradiated by a uniform EM plane wave, i.e.,

$$\hat{E}^{\mathrm{i}}_k(\boldsymbol{x},s) = \alpha_k\hat{e}^{\mathrm{i}}(s)\exp(-s\beta_m x_m/c_0), \quad (12)$$

where $\boldsymbol{\alpha}$ is the polarization vector, $\boldsymbol{\beta}$ is a unit vector in the direction of propagation and $\hat{e}^{\mathrm{i}}(s)$ is the amplitude. In the first example, we take a $x_3$-polarized plane wave propagating along the $x_1$-direction with a unit amplitude, i.e., $\boldsymbol{\alpha} = \boldsymbol{i}_3$, $\boldsymbol{\beta} = \boldsymbol{i}_1$ and $\hat{e}^{\mathrm{i}}(s) = 1.0\,\mathrm{V/m}$. We assume the time-harmonic time dependence, which corresponds to $s = \mathrm{i}\omega$ with $\omega = 2\pi f$, where $f$ is the operating frequency. The electric permittivity of Pyrex is taken to be $\epsilon_\mathrm{r} = \hat{\epsilon}_1/\epsilon_0 = 4.60 + \mathrm{i}0.023$ [5]. Fig. 2 shows the discretized computational model for $N = 10^3$ and the computed $E$-field distribution at $f = 12.60\,\mathrm{GHz}$. Note that the corresponding wavelength at this frequency is about $\lambda_1 \simeq 11.1\,\mathrm{mm}$. As a result, the chosen grid step $h = 1.0\,\mathrm{mm}$ is less than a tenth of the wavelength.

Before conducting the uncertainty analysis, we validate the implemented MoM-VIE model using an independent numerical method. For this purpose, we employ the finite integration technique (FIT) as implemented in CST Microwave Suite[1].

[1] Any mention of commercial products is for information only; it does not imply recommendation or endorsement by NIST.

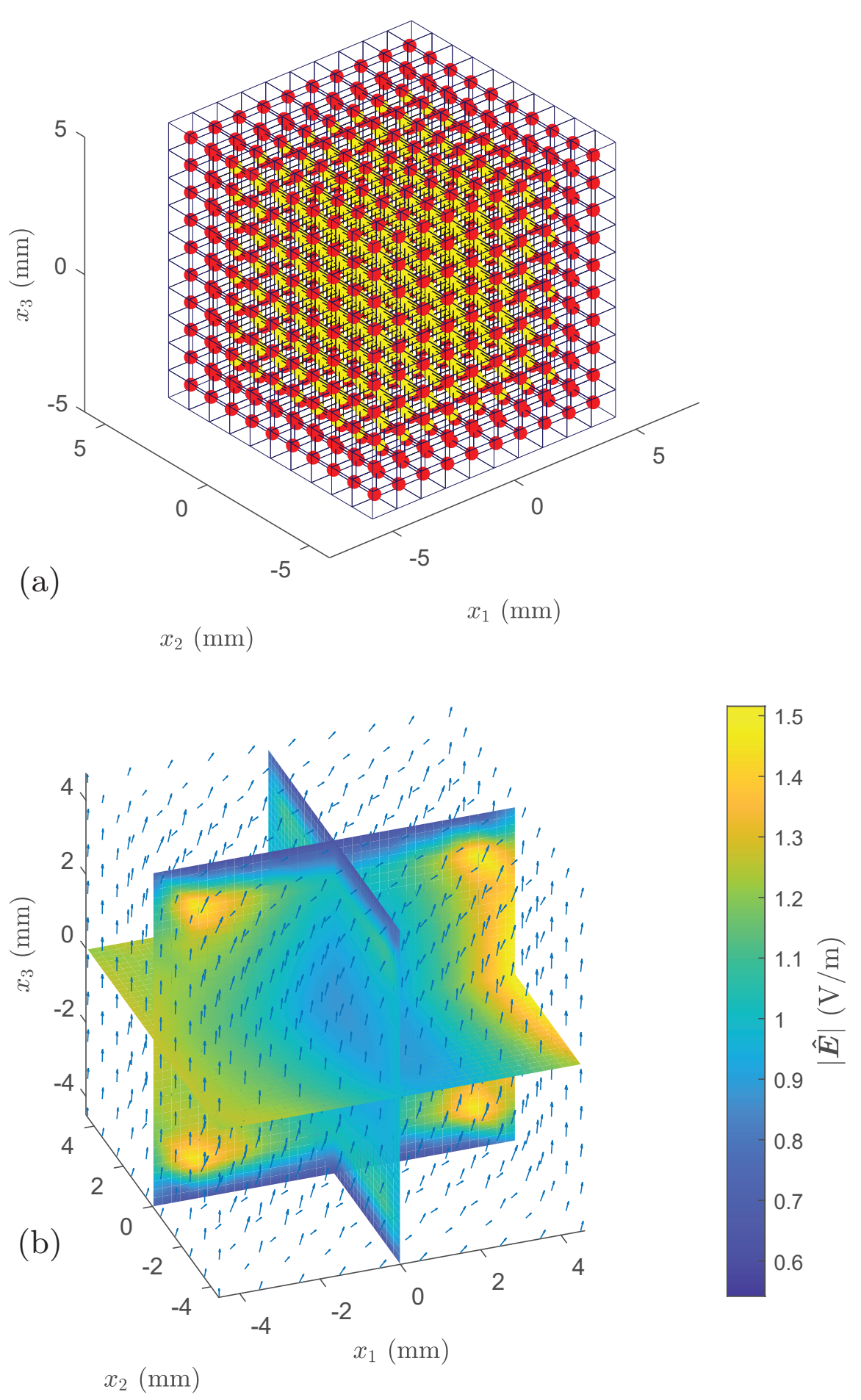

Fig. 2. The analyzed cubic cell. (a) VIE computational model; (b) Computed $E$-field distribution at $f = 12.60\,\text{GHz}$.

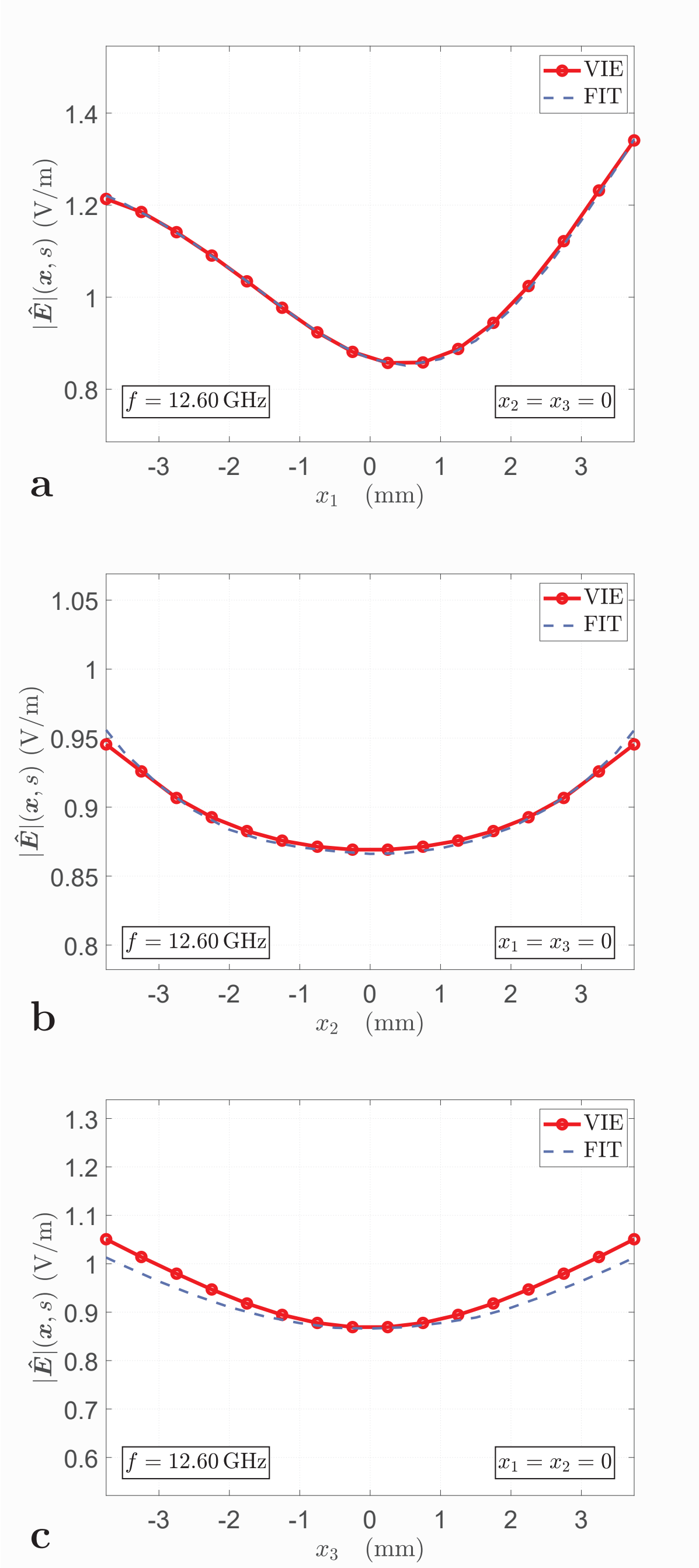

Fig. 3. The $E$-field distribution as calculated using MoM-VIE and FIT along (a) $x_1$-axis; (b) $x_2$-axis; (c) $x_3$-axis.

The MoM-VIE model was discretized into $N = 8 \cdot 10^3$ cubes, while the CST-FIT model consists of approximately $70 \cdot 10^3$ tetrahedrons. Fig. 3 presents the resulting total electric field distribution inside the cell evaluated along the three orthogonal axes intersecting at the origin. As shown, the results correlate very well.

### A. Type B Uncertainty Evaluation

The most straightforward way to estimate the uncertainty introduced by a vapor cell is through a *non-statistical* (Type B) uncertainty assessment (see Sec. VI). For the $E$-field distribution along the $x_2$-axis, this can be performed by calculating the rectangular-distribution uncertainty [17]

$$\Delta = \max\{\mathrm{M}[|\hat{\boldsymbol{E}}|_{\pm\delta}] - \mathrm{M}[|\hat{\boldsymbol{E}}|]\}/\sqrt{3}, \tag{13}$$

where $\mathrm{M}[|\hat{\boldsymbol{E}}|]$ represents the mean value of the $E$-field distribution along the central measurement line, while $\mathrm{M}[|\hat{\boldsymbol{E}}|]_{\pm\delta}$ are the mean values of the $E$-field measured along lines shifted slightly to either side of the center. The scaling factor $\sqrt{3}$ is the rectangular-distribution constant that transforms the worst-case shift into a statistical standard deviation. Fig. 4a shows the result of this analysis for the total $E$-field distribution at $f = 12.60\,\text{GHz}$ along the $x_2$-axis ($x_1 = x_3 = 0$) using an offset in the $x_1$-direction of $x_1 \pm \delta$, $\delta = 0.5\,\text{mm}$. Next, for each frequency in the chosen frequency band, the corresponding uncertainty $\Delta$ can be associated with the average electric field along the central measurement line. This procedure results in the plot presented in Fig. 4b. Among all computed uncertainties in $f \in \{5, 6, \cdots, 20\}\,\text{GHz}$, the (worst-case) maximum is $\Delta_{\text{max}} = 62.78\,\text{mV/m}$. A similar (Type B) uncertainty analysis can be carried out for an offset in the $x_3$-direction of $x_3 \pm \delta$. As shown in Fig. 5, misalignment along this axis results in a substantially lower uncertainty compared with the previous case. Indeed, the (worst-case) maximum is only $\Delta_{\max} = 2.16\,\mathrm{mV/m}$ for this configuration.

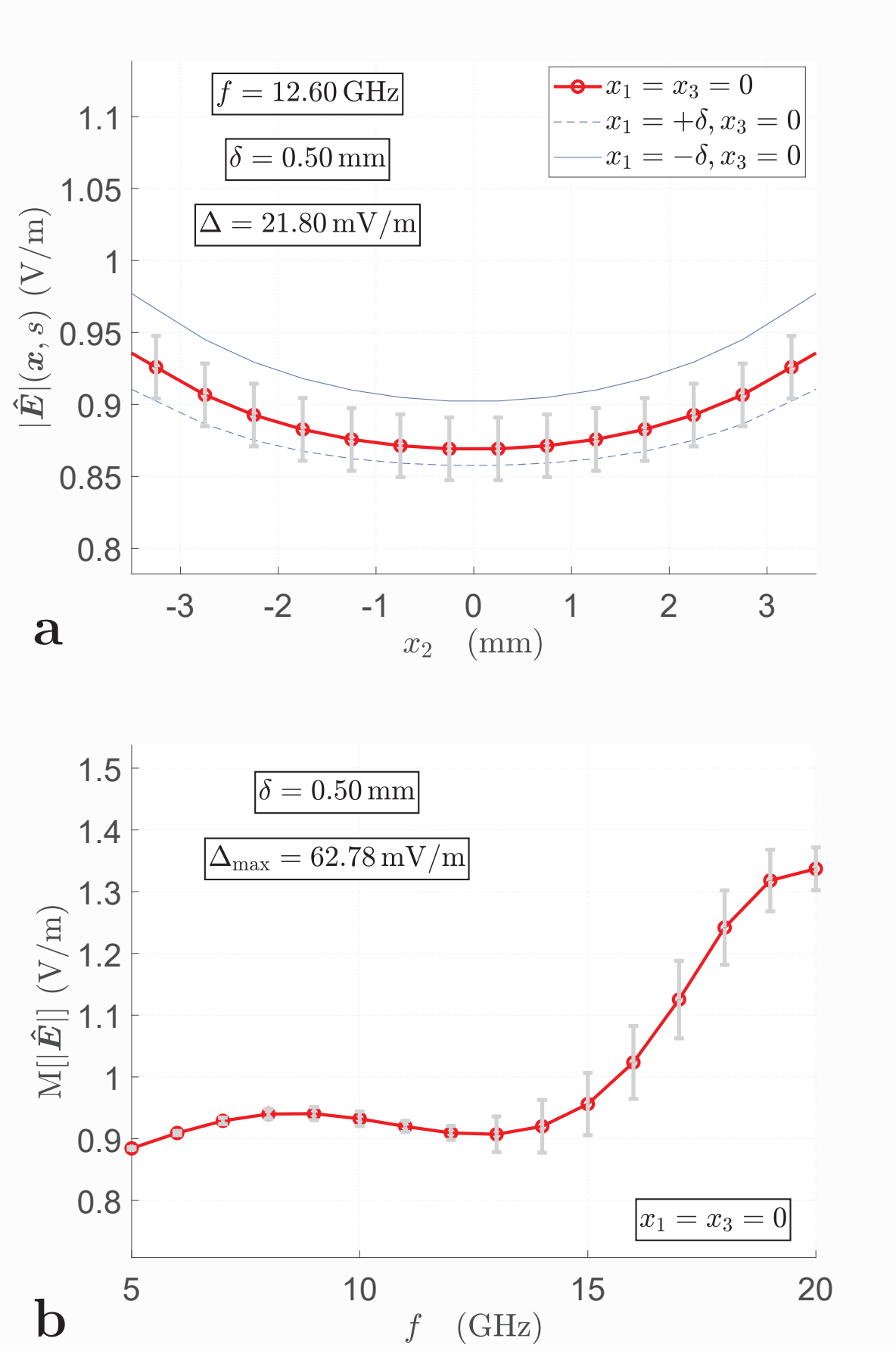

Fig. 4. Type B uncertainty derived from $x_1$-offset variations. (a) $E$-field distribution along the $x_2$-axis at $f = 12.60\,\mathrm{GHz}$; (b) Average $E$-field with respect to frequency.

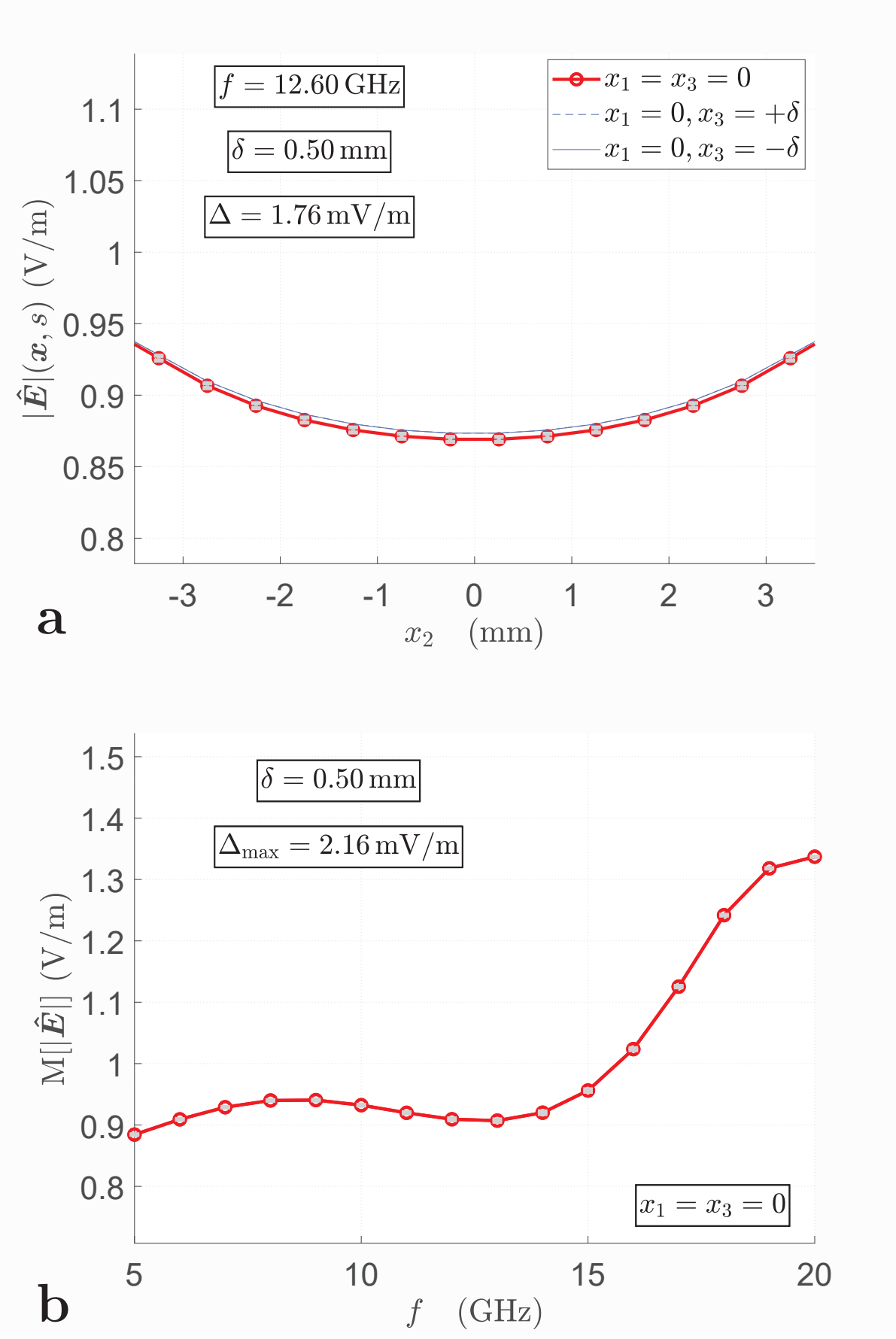

Fig. 5. Type B uncertainty derived from $x_3$-offset variations. (a) $E$-field distribution along the $x_2$-axis at $f = 12.60\,\mathrm{GHz}$; (b) Average $E$-field with respect to frequency.

### B. Type A Uncertainty Evaluation

The computational tool can also be applied to estimate *statistical* (Type A) uncertainty. For demonstration, we next examine how changes in material parameters affect the $E$-field distribution within the vapor cell. We assume that the (standard) deviation of $\mathrm{Re}(\epsilon_\mathrm{r})$ is $\mathrm{D}[\mathrm{Re}(\epsilon_\mathrm{r})] = 0.20$ around its mean value $\mathrm{M}[\mathrm{Re}(\epsilon_\mathrm{r})] = 4.60$. The other parameters remain the same as in the previous example. Fig. 6 shows the resulting mean values and 95% confidence intervals (CI) of the total $E$-field at grid points in $\mathcal{D}_0$ along the $x_1$-direction for $x_2 = x_3 = 0$ at $f = \{8.57, 12.60, 19.64\}\,\mathrm{GHz}$ (cf., [5]). In agreement with the results reported in [5], we observe that non-uniform standing-wave patterns become more pronounced at higher frequencies, where the cell's dimension is no longer relatively very small. We additionally investigated loss tangent uncertainty components with $\mathrm{M}[\mathrm{Im}(\epsilon_\mathrm{r})] = 0.023$ and $\mathrm{D}[\mathrm{Im}(\epsilon_\mathrm{r})] = 0.010$. Under these circumstances, the loss tangent uncertainty plays a minor role compared to $\delta\mathrm{Re}(\epsilon_r)$ with 95% confidence interval uncertainties of 0.05%, 0.21%, and 0.79% at $f$ =5 GHz, 10 GHz, and 20 GHz, respectively.

We further evaluate the uncertainty arising from variations in the plane-wave incidence angle. To do so, we write $\boldsymbol{\beta} = \cos(\phi)\boldsymbol{i}_1 + \sin(\phi)\boldsymbol{i}_2$ and assume the standard deviation $\mathrm{D}[\phi] = 5^\circ$ around the zero mean value $\mathrm{M}[\phi] = 0$, which corresponds to $\boldsymbol{\beta} = \boldsymbol{i}_1$. Fig. 7 shows the resulting $E$-field distributions along the $x_2$-axis as computed again for $f = \{8.57, 12.60, 19.64\}\,\mathrm{GHz}$. As can be seen, the lowest uncertainty in the $E$-field values along the $x_2$-axis (i.e., perpendicularly to the (mean) propagation direction $\boldsymbol{\beta} = \boldsymbol{i}_1$) occurs around the center of the cell.

Since we typically measure the average $E$-field across the cell in the direction perpendicular to the laser beams, we next evaluate the uncertainty of these average values over a chosen frequency range. Fig. 8 presents the results of the statistical analysis for variations in the cell's permittivity (with $\mathrm{M}[\mathrm{Re}(\epsilon_\mathrm{r})] = 4.60$, $\mathrm{D}[\mathrm{Re}(\epsilon_\mathrm{r})] = 0.20$) and in the direction of the plane-wave incidence (with $\mathrm{M}[\phi] = 0$, $\mathrm{D}[\phi] = 5^\circ$). As can be seen in Fig. 8, the variation in permittivity increases the uncertainty across most of the considered frequency range, $f \in [5, 20]\,\mathrm{GHz}$, whereas the misalignment of the plane-wave incidence direction becomes the source of uncertainty at only higher frequencies.

## VI. Discussion

In this paper, we examined several sources of uncertainty and assessed their impact on the measured $E$-field distribu-tion. The uncertainties, expressed relative to the mean value, are summarized in Table I. Here, $\delta_x$ and $\delta_z$ represent the spatial offsets in the $x$- and $z$-direction (see Figs. 4 and 5), respectively. Additional uncertainty components related to measuring the atomic spectrum have previously been reported [18]. They include calibrating the frequency axis, $\delta f_s$; identifying the atomic spectrum peaks $\delta f_c$; and calculations of the atomic dipole moment, $\delta p$. We include these components from Ref. [18] in Table I. Additionally, the measurement repeatability, $\delta s$, in Ref. [18] was measured to be 0.05%. We evaluated the combined uncertainty using the square root of the sum of squares [17, (1)]. Its values at selected frequencies are given in Table I and Fig. 9 shows the combined uncertainty in the form of error bars over $f \in \{5, 6, \cdots, 20\}$.

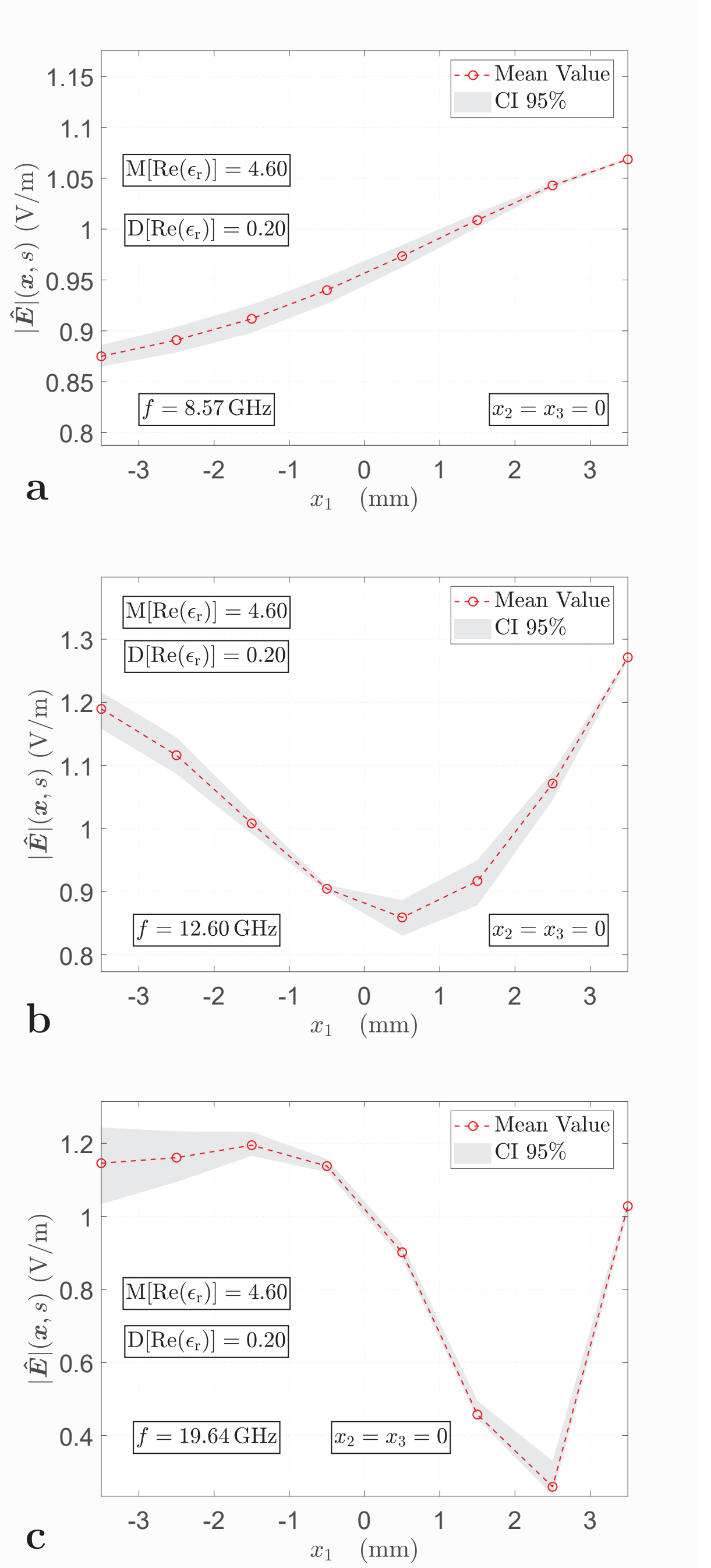

Fig. 6. Type A uncertainty of the $E$-field distribution caused by variations in the permittivity inside the cubic vapor cell along the $x_1$-axis at (a) $f = 8.57$ GHz; (b) $f = 12.60$ GHz; (c) $f = 19.64$ GHz.

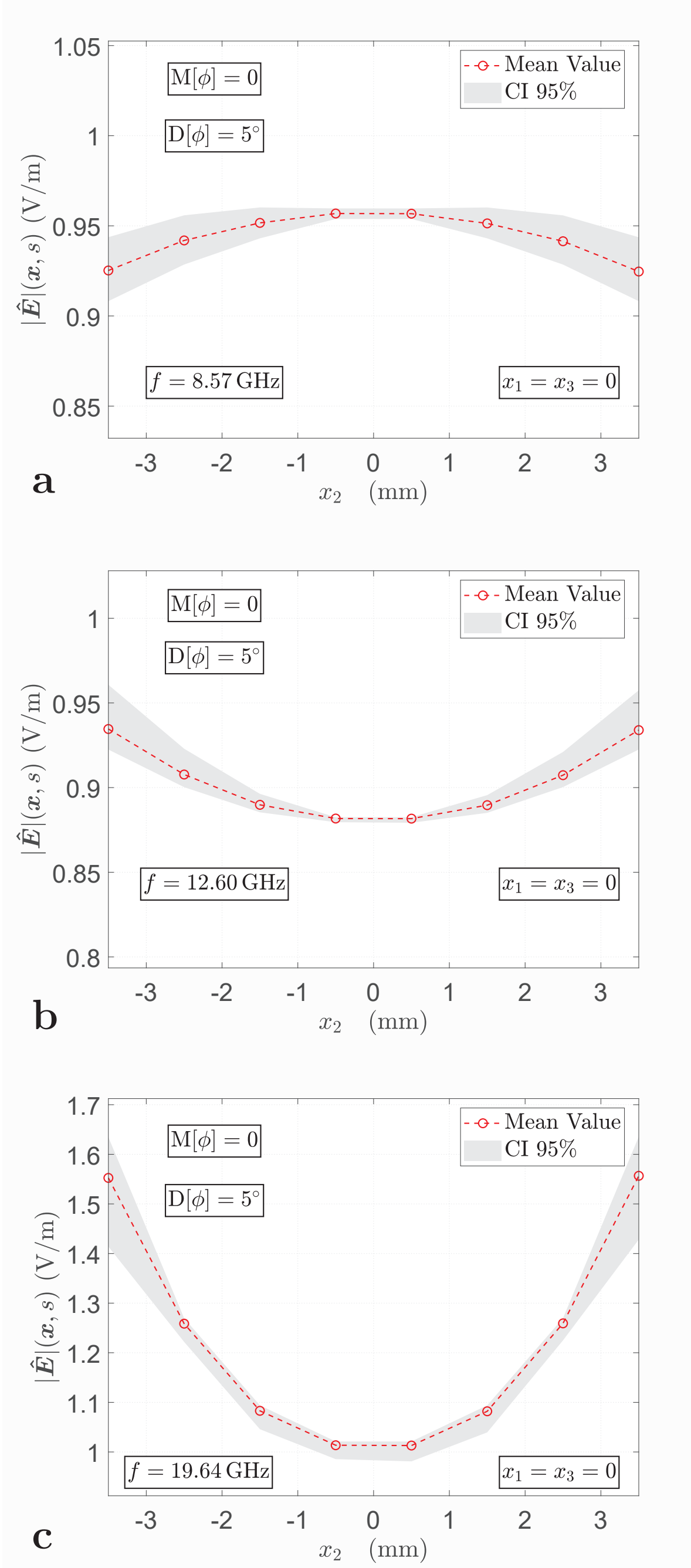

Fig. 7. Type A uncertainty of the $E$-field distribution caused by variations in the incidence angle inside the cubic vapor cell along the $x_2$-axis at (a) $f = 8.57$ GHz; (b) $f = 12.60$ GHz; (c) $f = 19.64$ GHz.

One limitation of our current study is we solely considered an idealized case in which the only dielectric component is the glass vapor cell. In reality, creating a portable Rydberg-based $E$-field probe requires additional dielectric components to couple in the required lasers for atomic excitation and readout [19]–[23]. Thus, the reported uncertainties should be considered a best case scenario.

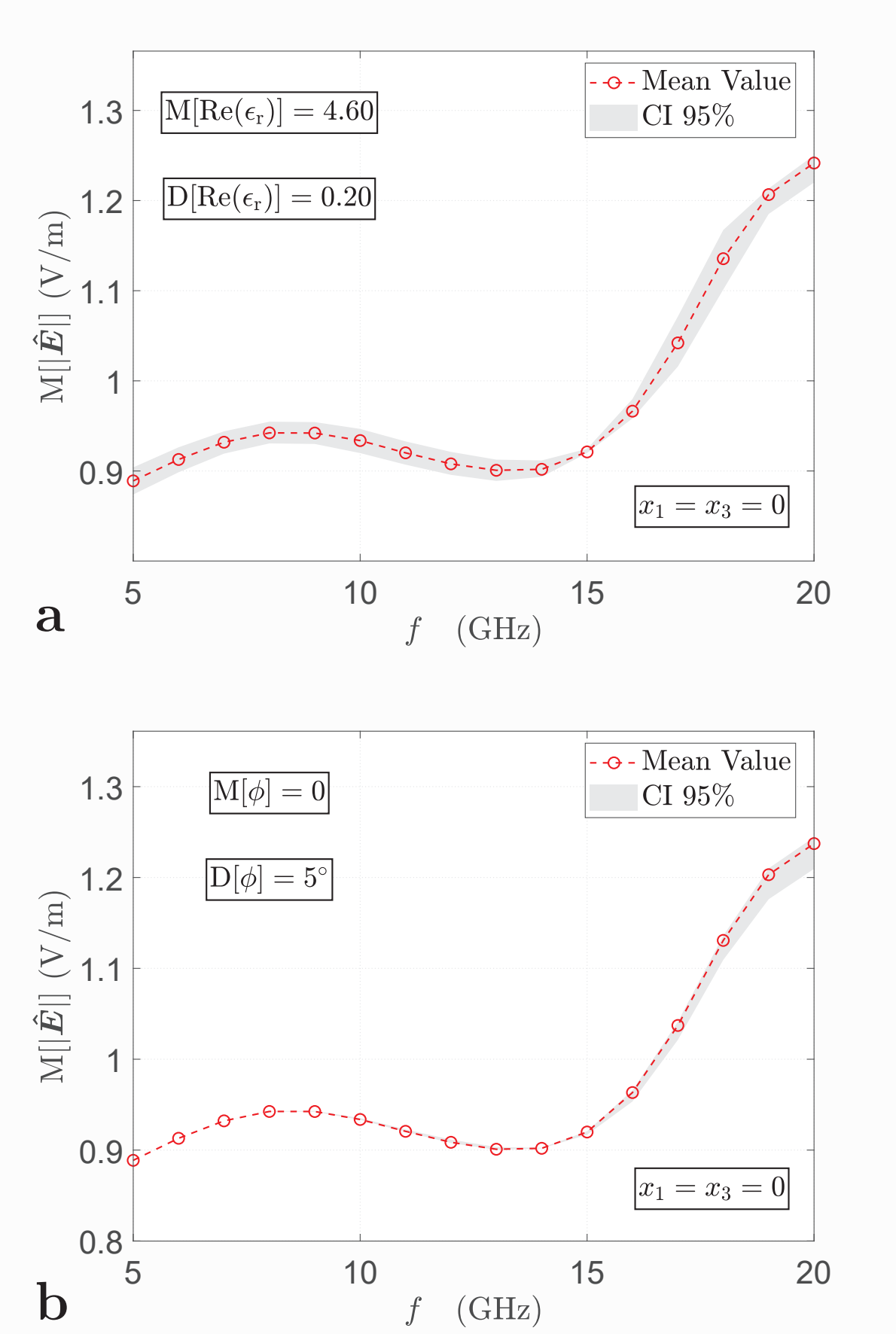

Fig. 8. Type A uncertainty of the average $E$-field along the $x_2$-axis inside the cubic vapor cell as caused by variations in (a) permittivity; (b) incidence angle.

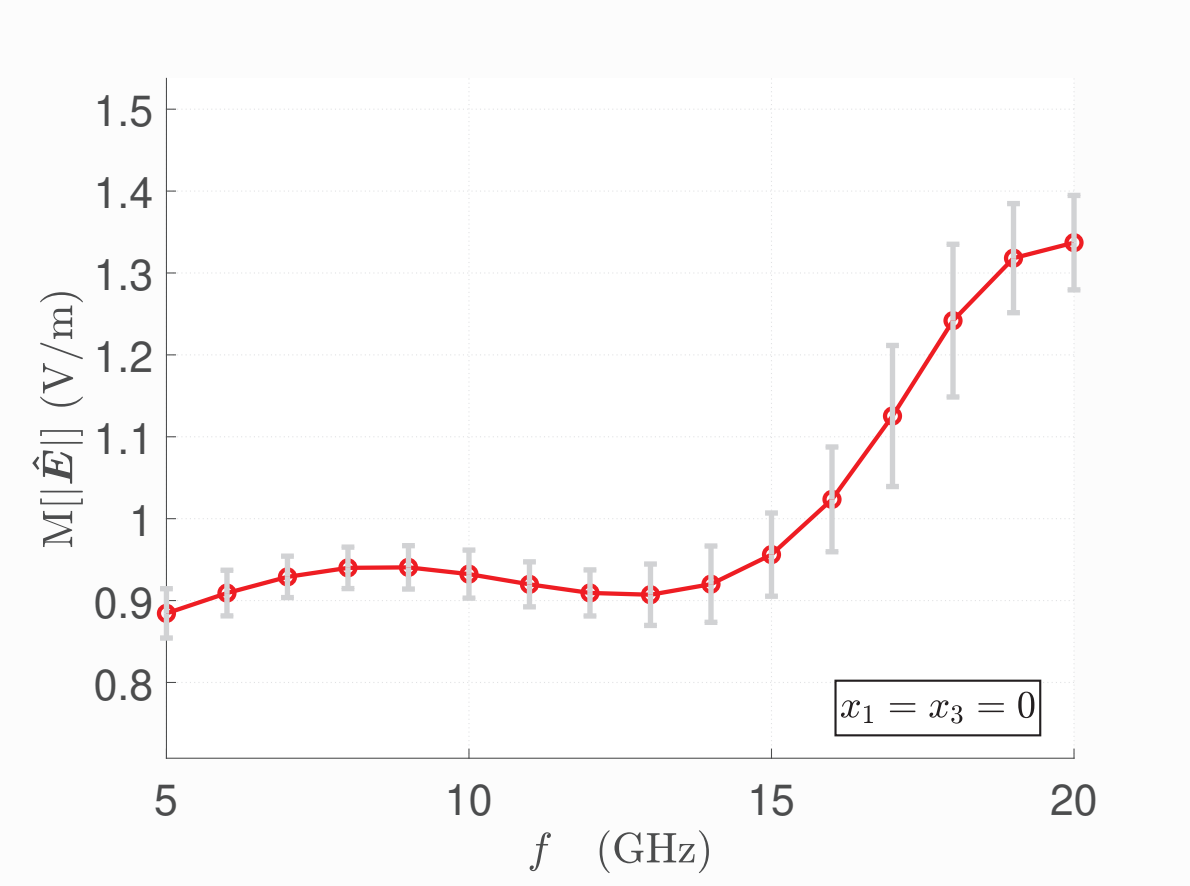

Fig. 9. Combined Type A and B uncertainty of the average $E$-field along the $x_2$-axis inside the cubic vapor cell

TABLE I
SUMMARY OF RELATIVE UNCERTAINTIES.

| Source | Type | $u_i(5\,\text{GHz})$ | $u_i(10\,\text{GHz})$ | $u_i(20\,\text{GHz})$ |
|---|---|---|---|---|
| $\delta\text{Re}(\epsilon_r)$ | A | 3.34% | 2.87% | 2.26% |
| $\delta\text{Im}(\epsilon_r)$ | A | 0.05% | 0.21% | 0.79% |
| $\delta\phi$ | A | 0.00% | 0.20% | 2.60% |
| $\delta_x$ | B | 0.20% | 1.27% | 2.60% |
| $\delta_z$ | B | 0.05% | 0.07% | 0.00% |
| $\delta f_s$ | B | 0.06% | 0.06% | 0.06% |
| $\delta f_c$ | B | 0.5% | 0.5% | 0.5% |
| $\delta p$ | B | 0.1% | 0.1% | 0.1% |
| $\delta s$ | A | 0.5% | 0.5% | 0.5% |
| $\Sigma$ | A & B | **3.42%** | **3.23%** | **4.45%** |

## VII. Conclusion

We presented an efficient numerical method for evaluating the EM scattering by vapor cells. The EM scattering problem was formulated through an electric-field VIE and subsequently solved with the aid of MoM. Special attention was given to the proper treatment of EM Green's tensor singularities.

This solution methodology establishes a controlled computational environment for analyzing EM scattering effects and statistical uncertainties arising from the unavoidable EM-field interaction with vapor cells. A comprehensive examination of the field measurement uncertainties was given.

## Appendix A
## EM Green's Tensor Analysis

A rigorous approach to addressing the strong singularity of EM Green's tensor has been presented in [15]. This method is based on excluding the "principal volume" from the integration domain over which the integration is expressed through the near-field (static) component of Green's tensor. The main advantage of this methodology is its flexibility in handling different shapes of the principal volume. The methodology introduced below is less flexible in this sense, but offers a straightforward way to evaluate both contribution of the (spherical) principal volume and the corresponding principal-value integral.

Thanks to the spatial shift invariance of the embedding, the Green's function can represented via a 3-D spatial Fourier representation [13, Sec. B.2]

$$\hat{G}(\boldsymbol{x}, s) = (2\pi)^{-3} \int_{\boldsymbol{k}\in\mathbb{R}^3} \exp(-\mathrm{i}\boldsymbol{k}\cdot\boldsymbol{x})\tilde{G}(\boldsymbol{k}, s)\mathrm{d}V(\boldsymbol{k}) \quad (14)$$

where $\boldsymbol{k}$ is the (real-valued) wave vector. Since the Fourier representation (14) transforms $\partial_m$ into $-\mathrm{i}k_m$, the corresponding transform-domain Green's tensor can be expressed via (5) as

$$\begin{aligned}\tilde{G}_{r,k}(\boldsymbol{k}, s) &= -(s^2/c_0^2)\tilde{g}(\boldsymbol{k}, s)\delta_{r,k} - k_r k_k \tilde{g}(\boldsymbol{k}, s) \\ &= -\frac{s^2/c_0^2}{k_m k_m + s^2/c_0^2}\delta_{r,k} - \frac{k_r k_k}{k_m k_m + s^2/c_0^2}. \end{aligned} \quad (15)$$

Before we evaluate the integration in $\boldsymbol{k}$-space, we shall rotate the coordinate reference frame $\{\boldsymbol{i}_1, \boldsymbol{i}_2, \boldsymbol{i}_3\}$ into a new one, say $\{\boldsymbol{i}'_1, \boldsymbol{i}'_2, \boldsymbol{i}'_3\}$, where the position vector $\boldsymbol{x} = r\cos(\phi)\sin(\theta)\boldsymbol{i}_1 + r\sin(\phi)\sin(\theta)\boldsymbol{i}_2 + r\cos(\theta)\boldsymbol{i}_3$ for $\{0 \leq r < \infty\}$, $\{0 \leq \phi < 2\pi\}$ and $\{0 \leq \theta \leq \pi\}$ is aligned with its polar axis along $\boldsymbol{i}_3'$. This rotation is accomplished by $R_{k,l}x_l$, where the rotation matrix is given by

$$\boldsymbol{R} = \begin{pmatrix} \cos(\phi)\cos(\theta) & \sin(\phi)\cos(\theta) & -\sin(\theta) \\ -\sin(\phi) & \cos(\phi) & 0 \\ \cos(\phi)\sin(\theta) & \sin(\phi)\sin(\theta) & \cos(\theta) \end{pmatrix}. \quad (16)$$

Expressing $\boldsymbol{k}$-vector in the new reference frame using spherical coordinates, say $\boldsymbol{k} = k\cos(\varphi)\sin(\vartheta)\boldsymbol{i}_1' + k\sin(\varphi)\sin(\vartheta)\boldsymbol{i}_2' + k\cos(\vartheta)\boldsymbol{i}_3'$ for $\{0 \le k < \infty\}$, $\{0 \le \varphi < 2\pi\}$ and $\{0 \le \vartheta \le \pi\}$, we get $k_m x_m = kr\cos(\vartheta)$ with $r = |\boldsymbol{x}|$, $k = |\boldsymbol{k}|$ and $\mathrm{d}V(\boldsymbol{k}) = k^2\sin(\vartheta)\mathrm{d}k\mathrm{d}\vartheta\mathrm{d}\varphi$. The rotation matrix (16) is orthogonal. As a result, under this transformation the Kronecker delta as well as $k_m k_m = k^2$ in (15) are invariant. Once this integration is carried out, the original $\hat{G}_{r,k}$ tensor can be found from the transformed one, say $\hat{G}'_{n,m}$, via

$$\hat{G}_{r,k} = R^{-1}_{r,n}\hat{G}'_{n,m}R^{-1}_{k,m} = R_{n,r}\hat{G}'_{n,m}R_{m,k}, \quad (17)$$

where we used the property that the inverse of a rotation matrix is equal to its transpose.

Hence, the use of (15) in (14) leads to

$$\hat{G}'_{p,q}(\boldsymbol{x},s) = \hat{M}'_{p,q}(\boldsymbol{x},s) + \hat{N}'_{p,q}(\boldsymbol{x},s), \quad (18)$$

where

$$\hat{M}'_{p,q}(\boldsymbol{x},s) = -(s^2/c_0^2)\hat{g}(\boldsymbol{x},s)\delta_{p,q}, \quad (19)$$

and

$$\hat{N}'_{p,q}(\boldsymbol{x},s) = -(2\pi)^{-3}\int_{k=0}^{\infty}\mathrm{d}k\frac{k^4}{k^2+s^2/c_0^2} \int_{\vartheta=0}^{\pi}\mathrm{d}\vartheta\sin(\vartheta)\exp[-\mathrm{i}kr\cos(\vartheta)]\int_{\varphi=0}^{2\pi}Q_{p,q}(\varphi,\vartheta)\mathrm{d}\varphi, \quad (20)$$

with $Q_{p,q} = k_p k_q/k^2$. The integration of $Q_{p,q}$ with respect to $\varphi$ is trivial and leads to non-vanishing diagonal terms $[\pi\sin^2(\vartheta)\delta_{1,q} + \pi\sin^2(\vartheta)\delta_{2,q} + 2\pi\cos^2(\vartheta)\delta_{3,q}]\delta_{p,q}$. Upon carrying out the integration with respect to $\vartheta$, we get

$$\hat{N}'_{1,1}(\boldsymbol{x},s) = \hat{N}'_{2,2}(\boldsymbol{x},s) = -(2\pi^2)^{-1}\int_{k=0}^{\infty}\frac{k^4}{k^2+s^2/c_0^2} \times\left[\frac{\sin(kr)}{k^3r^3} - \frac{\cos(kr)}{k^2r^2}\right]\mathrm{d}k, \quad (21)$$

$$\hat{N}'_{3,3}(\boldsymbol{x},s) = (2\pi^2)^{-1}\int_{k=0}^{\infty}\frac{k^4}{k^2+s^2/c_0^2} \times\left[\frac{2\sin(kr)}{k^3r^3} - \frac{2\cos(kr)}{k^2r^2} - \frac{\sin(kr)}{kr}\right]\mathrm{d}k. \quad (22)$$

Noting that the integrands in (21) and (22) are even functions of $k$, we further extend the range of integration for $k \in \mathbb{R}$. Subsequently, we analytically continue the integrands into the complex $k$-plane and evaluate the integrals with the aid of standard complex integration. To handle divergent integrals, we incorporate Dirac-delta distributions applying at $r = 0$. In this manner, upon collecting the results, we end up with

$$\hat{G}'_{1,1}(\boldsymbol{x},s) = \hat{G}'_{2,2}(\boldsymbol{x},s) = -\left(\frac{s^2}{c_0^2} + \frac{s/c_0}{r} + \frac{1}{r^2}\right)\hat{g}(\boldsymbol{x},s) + \frac{\delta(r)}{2\pi r^2}, \quad (23)$$

$$\hat{G}'_{3,3}(\boldsymbol{x},s) = 2\left(\frac{s/c_0}{r} + \frac{1}{r^2}\right)\hat{g}(\boldsymbol{x},s) - \frac{\delta(r)}{\pi r^2} + \frac{\delta'(r)}{2\pi r}. \quad (24)$$

Apparently, in contrast to $\hat{G}_{r,k}$ in the original reference frame, $\hat{G}'_{r,k}$ is diagonal and depends merely on the radial distance $r = |\boldsymbol{x}|$. Taking the former property into account, the transform relation (17) can be simplified to

$$\hat{G}_{r,k} = R_{1,r}\hat{G}'_{1,1}R_{1,k} + R_{2,r}\hat{G}'_{2,2}R_{2,k} + R_{3,r}\hat{G}'_{3,3}R_{3,k}. \quad (25)$$

Disregarding the Dirac-delta distributions at $r = 0$, the transformation (25) yields the same EM Green's tensor that could be obtained upon differentiating the scalar Green's function (6) for $r = |\boldsymbol{x} - \boldsymbol{x}'| \neq 0$ as indicated in (5).

Owing to their simplicity, the diagonal components of $G'_{r,k}$ as given by (23) and (24) can easily be integrated over a spherical domain. To that end, we divide a bounded spherical integration domain into two regions, say $\mathbb{B} = \mathbb{B}_\delta \cup \mathbb{B}_\Delta$, consisting of a small ball enclosing the origin, $\mathbb{B}_\delta = \{0 \le r < \delta, 0 \le \phi < 2\pi, 0 \le \theta \le \pi\}$ and $\mathbb{B}_\Delta = \{\delta \le r < R_0, 0 \le \phi < 2\pi, 0 \le \theta \le \pi\}$ with $\delta \downarrow 0$ and $R_0 > 0$. The Dirac-delta distributions operating at the origin yield non-vanishing contributions from $\mathbb{B}_\delta$, i.e.,

$$\int_{\mathbb{B}_\delta}\hat{G}'_{r,k}\mathrm{d}V = (\delta_{k,1} + \delta_{k,2} - 3\delta_{k,3})\delta_{r,k}, \quad (26)$$

while the integration over $\mathbb{B}_\Delta$ leads to

$$\int_{\mathbb{B}_\Delta}\hat{G}'_{r,k}\mathrm{d}V = (I\delta_{k,1} + I\delta_{k,2} + J\delta_{k,3})\delta_{r,k}. \quad (27)$$

Here, the diagonal terms are given by

$$\begin{aligned} I &= 2\exp(-sR_0/c_0) - 2\exp(-s\delta/c_0) \\ &\quad + (sR_0/c_0)\exp(-sR_0/c_0) - (s\delta/c_0)\exp(-s\delta/c_0) \\ &\quad + \mathrm{E}_1(sR_0/c_0) - \mathrm{E}_1(s\delta/c_0), \quad (28) \\ J &= -2\exp(-sR_0/c_0) + 2\exp(-s\delta/c_0) \\ &\quad - 2\mathrm{E}_1(sR_0/c_0) + 2\mathrm{E}_1(s\delta/c_0), \quad (29) \end{aligned}$$

where $\mathrm{E}_1(x)$ denotes the exponential integral [16, 5.1.4]. Since the trace of the spherical-volume integrals is invariant to the rotation (17), i.e.,

$$\mathrm{Tr}\int_{\mathbb{B}}\hat{G}_{r,k}\mathrm{d}V = \mathrm{Tr}\int_{\mathbb{B}}\hat{G}'_{r,k}\mathrm{d}V, \quad (30)$$

we can use (26) and (27) to evaluate the right-hand side of (30), which then yields

$$\int_{\mathbb{B}_\delta}\hat{G}_{r,k}\mathrm{d}V = -\frac{1}{3}\delta_{r,k}, \quad (31)$$

$$\int_{\mathbb{B}_\Delta}\hat{G}_{r,k}\mathrm{d}V = \frac{2}{3}\left[\left(1 + \frac{sR_0}{c_0}\right)\exp\left(-\frac{sR_0}{c_0}\right) - 1\right]\delta_{r,k}, \quad (32)$$

where we used the spherical symmetry in the original coordinate system. Apparently, (31) corresponds to the contribution from the spherical principal volume $\mathbb{B}_\delta$ (cf., (7)), while (32) can be interpreted as the principal-value integral (cf., (9) and [10, Appendix]).

## ACKNOWLEDGMENT

M. Stumpf acknowledges support from the Fulbright Scholar Program, sponsored by the U.S. Department of State and the J. William Fulbright Commission in the Czech Republic. The views expressed are solely those of the author and do not represent the official views of the U.S. government, the Fulbright Program, or the Commission.

## REFERENCES

[1] C. L. Holloway, J. A. Gordon, S. Jefferts, A. Schwarzkopf, D. A. Anderson, S. A. Miller, N. Thaicharoen, and G. Raithel, "Broadband rydberg atom-based electric-field probe for SI-traceable, self-calibrated measurements," *IEEE Trans. Antennas Propag.*, vol. 62, no. 12, pp. 6169–6182, 2014.

[2] H. Fan, S. Kumar, J. Sedlacek, H. Kübler, S. Karimkashi, and J. P. Shaffer, "Atom based RF electric field sensing," *J. Phys. B: At. Mol. Opt. Phys.*, vol. 48, no. 20, p. 202001, 2015.

[3] C. L. Holloway, M. T. Simons, J. A. Gordon, A. Dienstfrey, D. A. Anderson, and G. Raithel, "Electric field metrology for SI traceability: Systematic measurement uncertainties in electromagnetically induced transparency in atomic vapor," *J. Appl. Phys.*, vol. 121, no. 23, 2017.

[4] N. Schlossberger, N. Prajapati, S. Berweger, A. P. Rotunno, A. B. Artusio-Glimpse, M. T. Simons, A. A. Sheikh, E. B. Norrgard, S. P. Eckel, and C. L. Holloway, "Rydberg states of alkali atoms in atomic vapour as SI-traceable field probes and communications receivers," *Nature Reviews Physics*, vol. 6, no. 10, pp. 606–620, Oct 2024. [Online]. Available: https://doi.org/10.1038/s42254-024-00756-7

[5] H. Fan, S. Kumar, J. Sheng, J. P. Shaffer, C. L. Holloway, and J. A. Gordon, "Effect of vapor-cell geometry on Rydberg-atom-based measurements of radio-frequency electric fields," *Phys. Rev. Appl.*, vol. 4, no. 4, p. 044015, 2015.

[6] L. Zhang, Z. Li, S. Liu, S. Xu, J. Kong, R. Zhao, H. Guo, H. Wen, X. Li, Z. Ma *et al.*, "Influence of the size of the cubic atomic vapor cell on a Rydberg atomic microwave sensor," *Applied Optics*, vol. 63, no. 34, pp. 8802–8807, 2024.

[7] D. Maurya, A. Artusio-Glimpse, A. Meraki, D. Lopez, and V. Aksyuk, "Optimizing vapor cells for Rydberg atom-based electrometer applications," *arXiv preprint arXiv:2509.07823*, 2025.

[8] A. Naghibi, F. D. Nayeri, R. Azimirad, M. Davoudi-Darareh, and S. Poursajadi, "Electromagnetic analysis and resonance-based optimization of atomic vapor cells for enhanced accuracy and sensitivity in Rydberg RF sensors," *IEEE Sens. J.*, 2026.

[9] R. F. Harrington, *Field Computation by Moment Methods*. Piscataway, NJ: Wiley–IEEE Press, 1993.

[10] D. E. Livesay and K.-M. Chen, "Electromagnetic fields induced inside arbitrarily shaped biological bodies," *IEEE Trans. Microw. Theory Tech.*, vol. 22, no. 12, pp. 1273–1280, 1974.

[11] A. Lakhtakia and G. W. Mulholland, "On two numerical techniques for light scattering by dielectric agglomerated structures," *J. Res. Natl. Inst. Stand. Technol.*, vol. 98, no. 6, p. 699, 1993.

[12] Joint Committee for Guides in Metrology, Working Group 1, "Evaluation of measurement data – guide to the expression of uncertainty in measurement," *Int. Organ. Stand. Geneva ISBN*, vol. 50, p. 134, 2008.

[13] A. T. de Hoop, *Handbook of Radiation and Scattering of Waves*. London, UK: Academic Press, 1995.

[14] J. Van Bladel, "Some remarks on Green's dyadic for infinite space," *IEEE Trans. Antennas Propag.*, vol. 9, no. 6, pp. 563–566, 1961.

[15] A. D. Yaghjian, "Electric dyadic Green's functions in the source region," *Proc. IEEE*, vol. 68, no. 2, pp. 248–263, 1980.

[16] M. Abramowitz and I. A. Stegun, *Handbook of Mathematical Functions*. New York, NY: Dover Publications, 1972.

[17] D. A. Hill and M. Kanda, "Measurement uncertainty of radiated emissions," National Institute of Standards and Technology (NIST), Boulder, CO, Technical Note 1389, March 1997. [Online]. Available: https://nvlpubs.nist.gov/nistpubs/Legacy/TN/nbstechnicalnote1389.pdf

[18] M. T. Simons, M. D. Kautz, J. A. Gordon, and C. L. Holloway, "Uncertainties in rydberg atom-based rf e-field measurements," in *2018 International Symposium on Electromagnetic Compatibility (EMC EUROPE)*. IEEE, 2018, pp. 376–380.

[19] M. T. Simons, J. A. Gordon, and C. L. Holloway, "Fiber-coupled vapor cell for a portable Rydberg atom-based radio frequency electric field sensor," *Applied Optics*, vol. 57, no. 22, pp. 6456–6460, 2018.

[20] D. A. Anderson, R. E. Sapiro, and G. Raithel, "A self-calibrated SI-traceable Rydberg atom-based radio frequency electric field probe and measurement instrument," *IEEE Trans. Antennas Propag.*, vol. 69, no. 9, pp. 5931–5941, 2021.

[21] R. Mao, Y. Lin, K. Yang, Q. An, and Y. Fu, "A high-efficiency fiber-coupled Rydberg-atom integrated probe and its imaging applications," *IEEE Antennas Wireless Propag. Lett.*, vol. 22, no. 2, pp. 352–356, 2022.

[22] R. Mao, Y. Lin, Y. Fu, Y. Ma, and K. Yang, "Digital beamforming and receiving array research based on Rydberg field probes," *IEEE Trans. Antennas Propag.*, vol. 72, no. 2, pp. 2025–2029, 2023.

[23] W. J. Watterson, N. Prajapati, R. Castillo-Garza, S. Berweger, N. Schlossberger, A. Artusio-Glimpse, C. L. Holloway, and M. T. Simons, "An imaging radar using a Rydberg atom receiver," *Applied Physics Letters*, vol. 127, no. 16, 2025.